\documentstyle[11pt,newpasp,twoside,psfig]{article}
\index{pulsars!Vela}
\index{pulsar wind nebulae!G263.9-3.3 }
\index{synchrotron emission}
\index{Chandra}

\def\cxo{{\em Chandra}}

\def\lesssim{\mathrel{\hbox{\rlap{\hbox{\lower4pt\hbox{$\sim$}}}\hbox{$<$}}}}
\def\gtrsim{\mathrel{\hbox{\rlap{\hbox{\lower4pt\hbox{$\sim$}}}\hbox{$>$}}}}
\def\edcomment#1{\iffalse\marginpar{\raggedright\sl#1\/}\else\relax\fi}
\reversemarginpar

\begin{document}

\title{Spatially resolved spectrum of the Vela PWN}
 \author{Oleg Kargaltsev and George Pavlov}
\affil{Dept.\ of Astronomy and Astrophysics, The Pennsylvania
State University, 525 Davey Lab, University Park, PA 16802;
green@astro.psu.edu, pavlov@astro.psu.edu}

\begin{abstract}

A series of 13 \cxo\ observations provided the deepest images of
the Vela PWN yet available.
In addition to the fine structure of the inner PWN features,
a much larger and fainter asymmetric X-ray nebula emerges in the
summed images. The shape of this outer PWN is similar to that of
the radio PWN. We also present the spectral map of the Vela PWN
that reveals a shell of soft emission surrounding the inner PWN
and an extended region of harder emission south-west of the
pulsar. This may indicate that the outer jet is supplying
particles to this region.

\end{abstract}

\section{Introduction}

 \indent Observations of the Vela and Crab pulsar-wind nebulae
 (PWNe) with the {\sl Chandra X-ray Observatory} have
 revealed the complex structure and dynamics of relativistic outflows from pulsars
 (e.g., Hester et al.\ 2002; Pavlov et al.\ 2003).
 The innermost brightest parts of these PWNe exhibit approximately
axially-symmetric morphologies, with extended jet-like structures
stretched along the symmetry axis (Helfand et al.\ 2001; Pavlov et
al.\ 2001).

To better understand the properties of the shocked pulsar wind, it
is important to investigate the correlation between the spectral
and spatial structures of the PWN. For instance, {\sl XMM}
observations of the Crab PWN (with $5''$ resolution) have shown
that the spatial dependence of the spectral slope is not
isotropic, being well correlated with the PWN structure: e.g., the
hardest emission comes from the inner torus region (Willingale et
al.\ 2001). The \cxo\ resolution and proximity of the Vela pulsar
($d\simeq 300$ pc; Dodson et al.\ 2003a) make it possible to
obtain an even better quality spectral map of the Vela PWN,
provided that a sufficient number of counts is collected.

Here we present the spectral map of the Vela PWN obtained from
deep observations with the \cxo\ ACIS detector.
 We also
investigate the large-scale morphology of the Vela PWN in X-rays
and compare it with that of the radio PWN  (Dodson et al.\ 2003b).

\section{X-rays: PWN spatial and spectral structure}

\indent The large-scale X-ray structure of the Vela PWN is shown
in the summed images composed of eight ACIS and three HRC
observations (left and right panels of Fig.~1\footnote{see {\tt
http://www.astro.psu.edu/users/green/IAU2003/ for the color images
}}, respectively; see Pavlov et al.\ 2003 for technical details of
 image reduction). The brightness scale in the images is adjusted to
 emphasize
a fainter, more diffused emission (the inner PWN features are
shown by contours in the left panel of Fig.~1; see also Fig.~3).
This faint emission is clearly asymmetric: most of it is located
south-west (SW) of the symmetry axis that is approximately
co-aligned with the direction of the pulsar's proper motion (left
panel in Fig.~1). Such asymmetry, as well as persistent bending of
the {\sl northeastern jet} to SW, suggest the presence of
large-scale ($\gtrsim 1$ pc) density/pressure gradients or a wind
(invisible in X-rays) blowing in the SW direction in the ambient
SNR medium. Another region of even fainter emission is seen at the
bottom of the image shown in the right panel of Figure 1. This
emission could be produced by the pulsar wind at the time when the
pulsar was located close to its birth place. If true, this places
an upper limit of $\approx2(t_{\rm age}/11,000~{\rm
yrs})^{-2/3}(E/1~{\rm keV})^{-1/3} $ $\mu$G on the magnetic field
in the region. Alternatively, the emission could be due to a
background SNR filament. To obtain the spectrum and resolve this
issue, a deep exposure of this region with ACIS is needed.

\begin{figure}[hh] \begin{center}
\vspace{-0.2cm}
 \hbox{
\psfig{figure=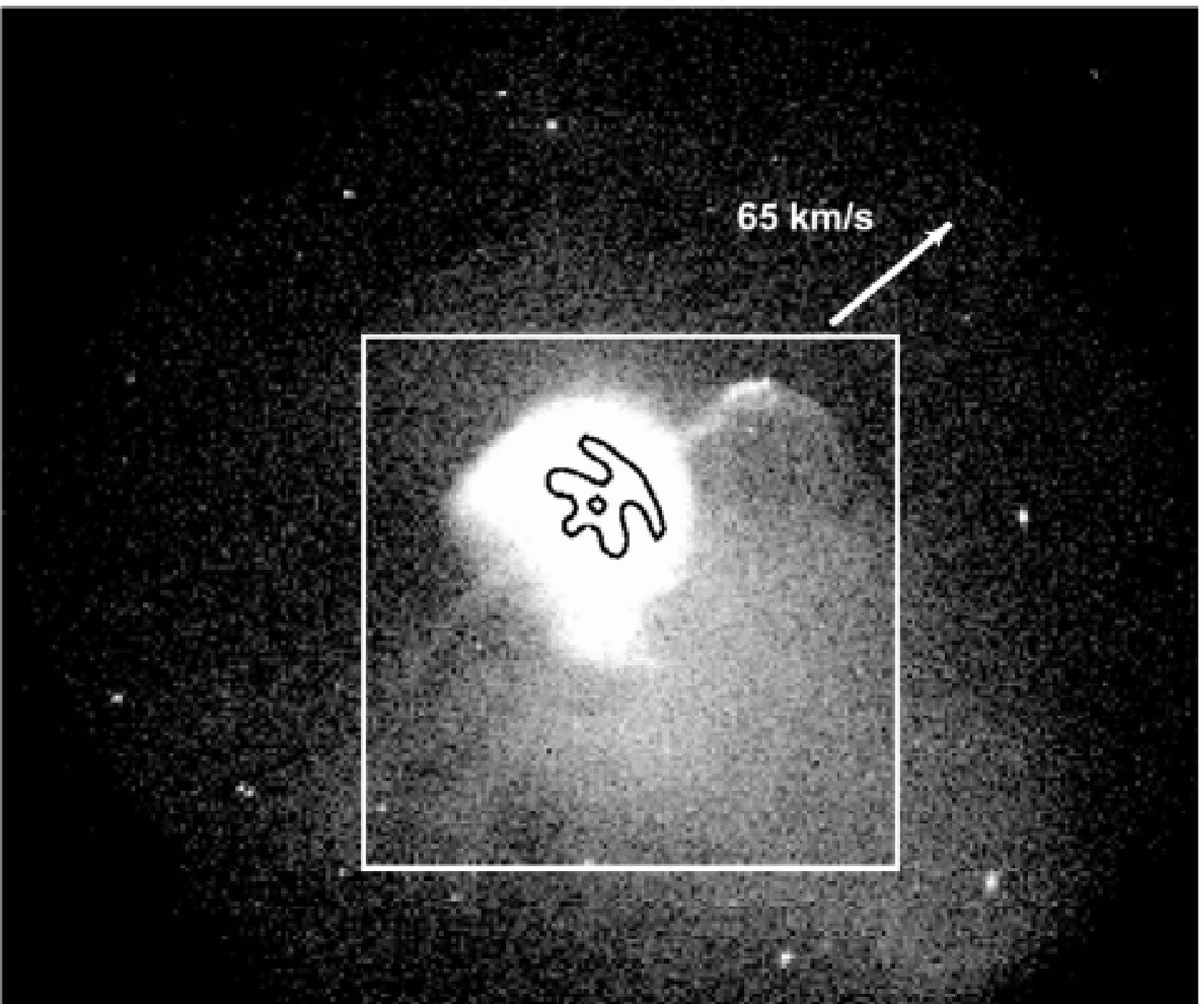,height=6.0cm} \hspace{0.1cm}
\psfig{figure=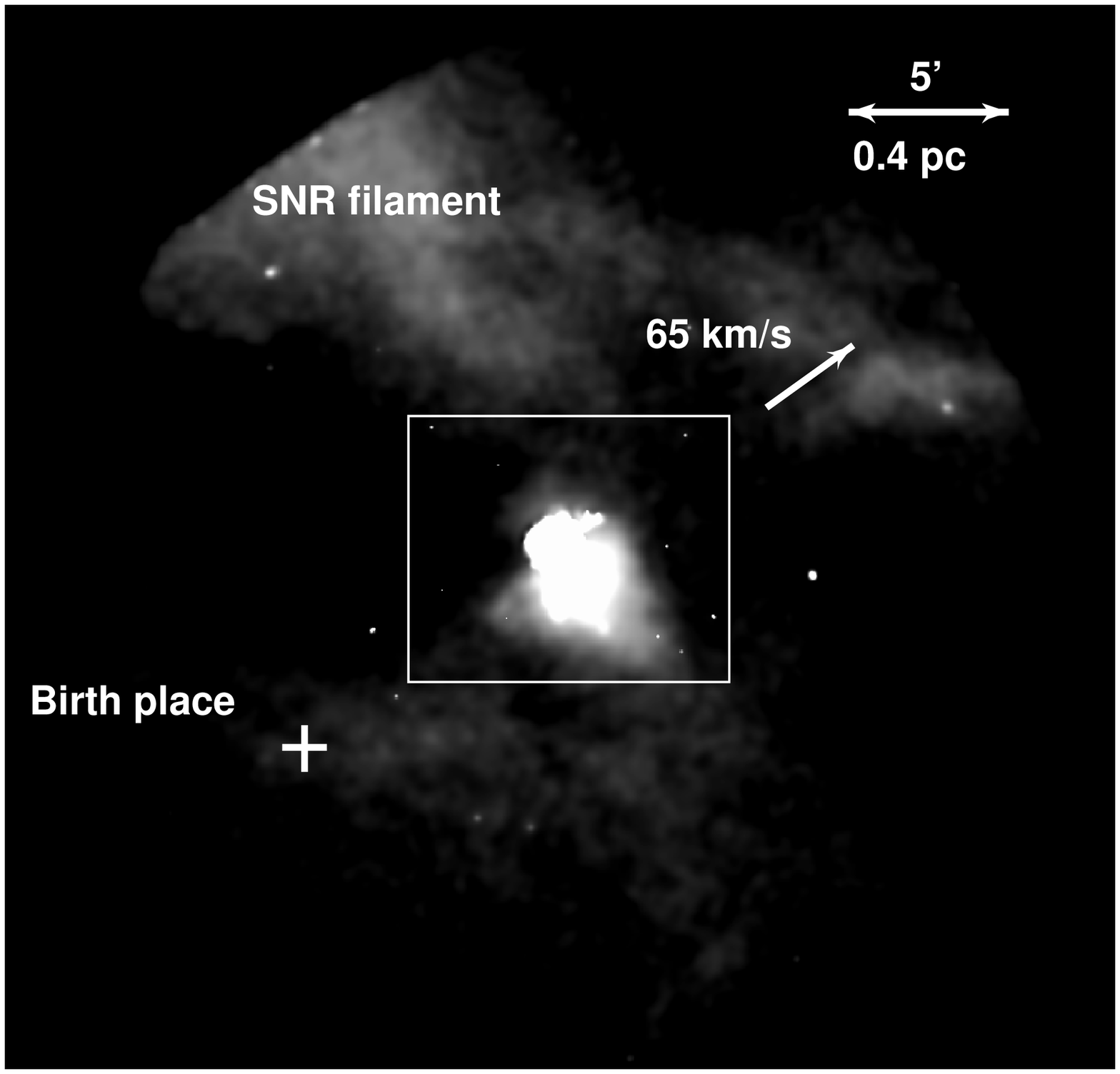,height=6.0cm}
 }
\vspace{-0.5cm} \end{center}\caption{ {\bf Left:} Summed ACIS-S
image of the Vela PWN (size $9\farcm4 \times 7\farcm8$; exposure
time $\approx 160$ ks). The black contours show the features of
the inner PWN which is saturated in this image. The white box
shows the region used
 to build the spectral map (Fig.~2). {\bf Right:} Vela PWN
image obtained by combining 3 HRC observations (size $35' \times
34'$; exposure time $\approx 150$ ks). The inner box shows the
size of the region displayed in the left panel.}
 \label{fig1} \end{figure}

To investigate the PWN spectral properties, we produced a photon
index map of a $4\farcm2 \times 4\farcm2$ region (white box in the
left panel of Fig.~1) from the
 merged ACIS event list with the aid of CIAO software\footnote{Chandra
Interactive Analysis of Observations
--- see {\tt http://cxc.harvard.edu/ciao/}}, v.2.3.
Although extracting the spectra from the merged event list is
generally not recommended\footnote{ {\tt
http://cxc.harvard.edu/ciao2.3/threads/combine/}}, we believe that
in this case the uncertainties introduced by doing so are minimal
since the PWN was imaged at the same location on the ACIS S3 chip
in all 8 observations, and a properly averaged\footnote{ {\tt
http://cxc.harvard.edu/ciao2.3/threads/wresp\_multiple\_sources/index.html\#warf}}
 ARF was used. After some experimenting, we adopted $2\farcs5$ and
$10''$ bins to be used for spectral extraction, depending on the
distance from the pulsar. This provides roughly comparable
signal-to-noise ratios for the bright inner PWN and much fainter
outer PWN.

 Most noticeable features in the spectral
map (Fig.~2, left panel) are the {\em shell of soft emission}
surrounding the inner PWN (which includes the arcs and the inner
jets with hard spectra; Fig.~2, right panel) and the relatively
{\em hard spectrum of the diffuse emission SW of the pulsar}
(which is harder, on average, than that of the shell located
closer to the pulsar). The harder spectrum can be explained if the
bent outer jet (see Fig.~1), whose spectrum is also hard (photon
index $\Gamma\approx 1.3$), supplies particles to the region SW of
the pulsar on a timescale shorter than the synchrotron cooling
time $t_{\rm syn}\approx 39 (B/100~{\rm \mu G})^{-3/2}(E/1~{\rm
keV})^{-1/2}$ years. The energy losses for the particles carried
with the equatorial outflow can be larger than for the jet's
particles, e.g., due to additional expansion losses which are less
important for the well-collimated jet. Emission produced by the
particles associated with the equatorial outflow can give rise to
the {\em soft shell} if the particles move outwards sufficiently
slow and have enough time to cool.

 \begin{figure}[hh] \begin{center}
 \vspace{-0.4cm}
 \hbox{
\psfig{figure=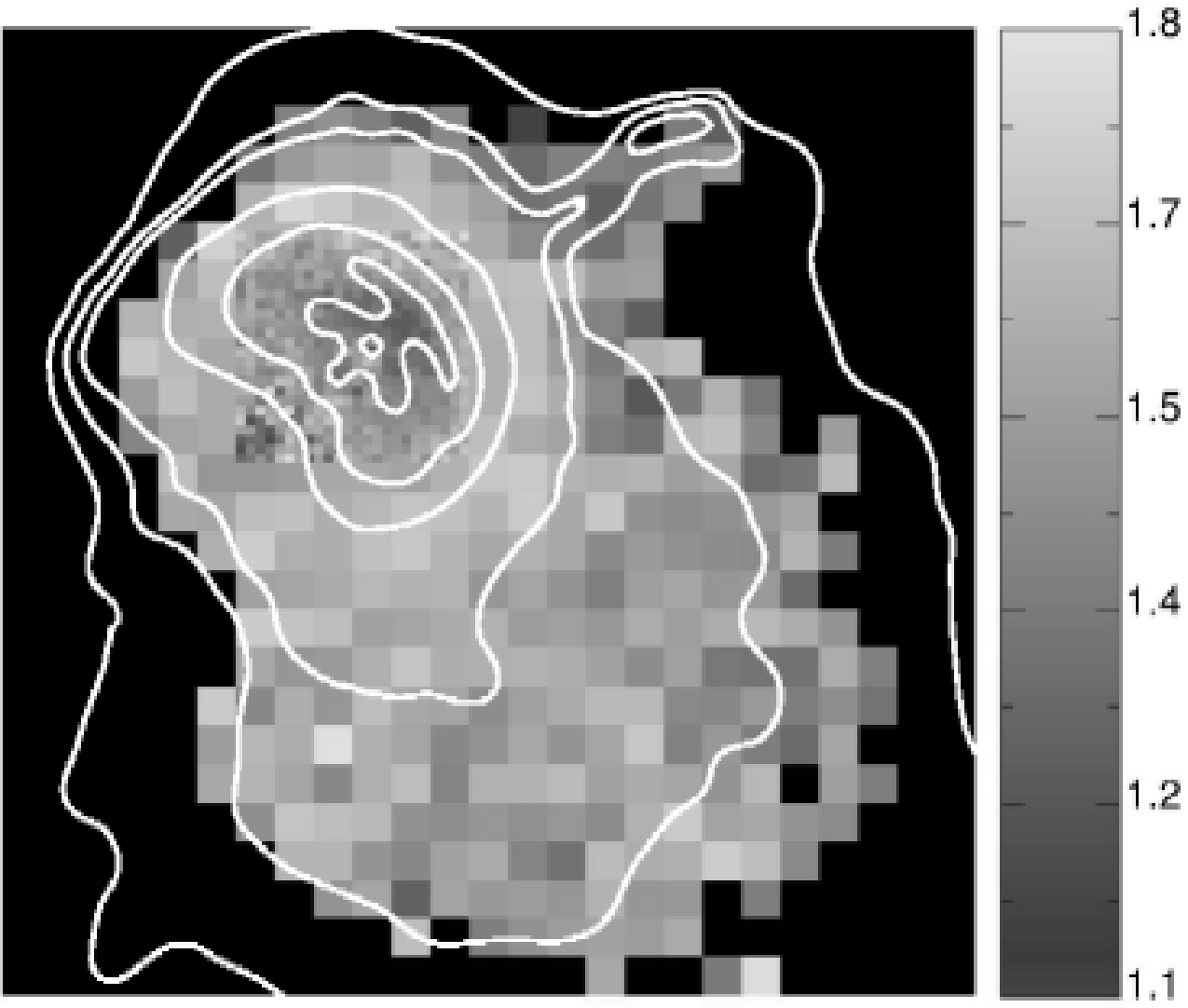,height=6.0cm} \hspace{0.1cm}
\psfig{figure=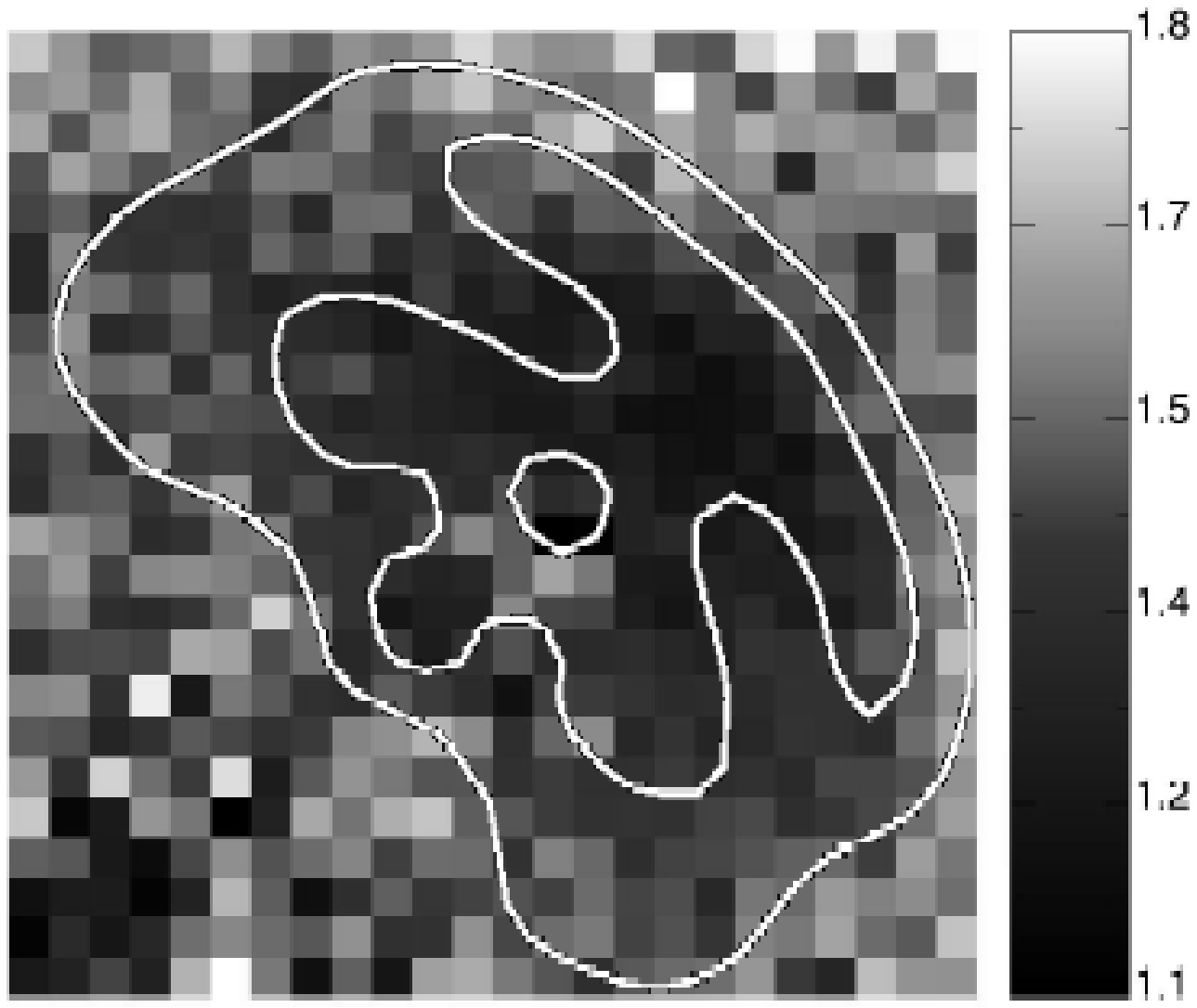,height=6.0cm}
 }
\vspace{-0.8cm} \end{center}\caption{ Photon index maps
($4\farcm2\times 4\farcm2$ - left; $1' \times 1'$ - right) with
X-ray contours overlayed. In the left panel, black color
corresponds to the pixels with low S/N ratio for which spectral
index is not calculated.}

 \label{fig3} \end{figure}

\section{Connection between the radio and X-rays}
In the recently obtained radio images of the Vela PWN the emission
mostly comes from two lobes of different sizes and brightnesses
(Dodson et al. 2003b). Overlaying the X-ray contours on top of the
radio image, we find that the outer contours (corresponding to
lower X-ray brightness) are well correlated with the shape of the
radio PWN (Fig.~3). This suggests that the X-ray and radio
emitting electrons are carried with the same outflow which is
mostly confined to a low-latitude (equatorial) region. In terms of
relative brightness, the radio emission is brighter further away
from the pulsar while the X-ray emission is the brightest close to
the pulsar. Such a picture can be explained by the synchrotron and
expansion losses which particles suffer as they travel away from
the pulsar.

\begin{figure}[hh] \begin{center}
\vspace{-0.3cm}
 \hbox{
\psfig{figure=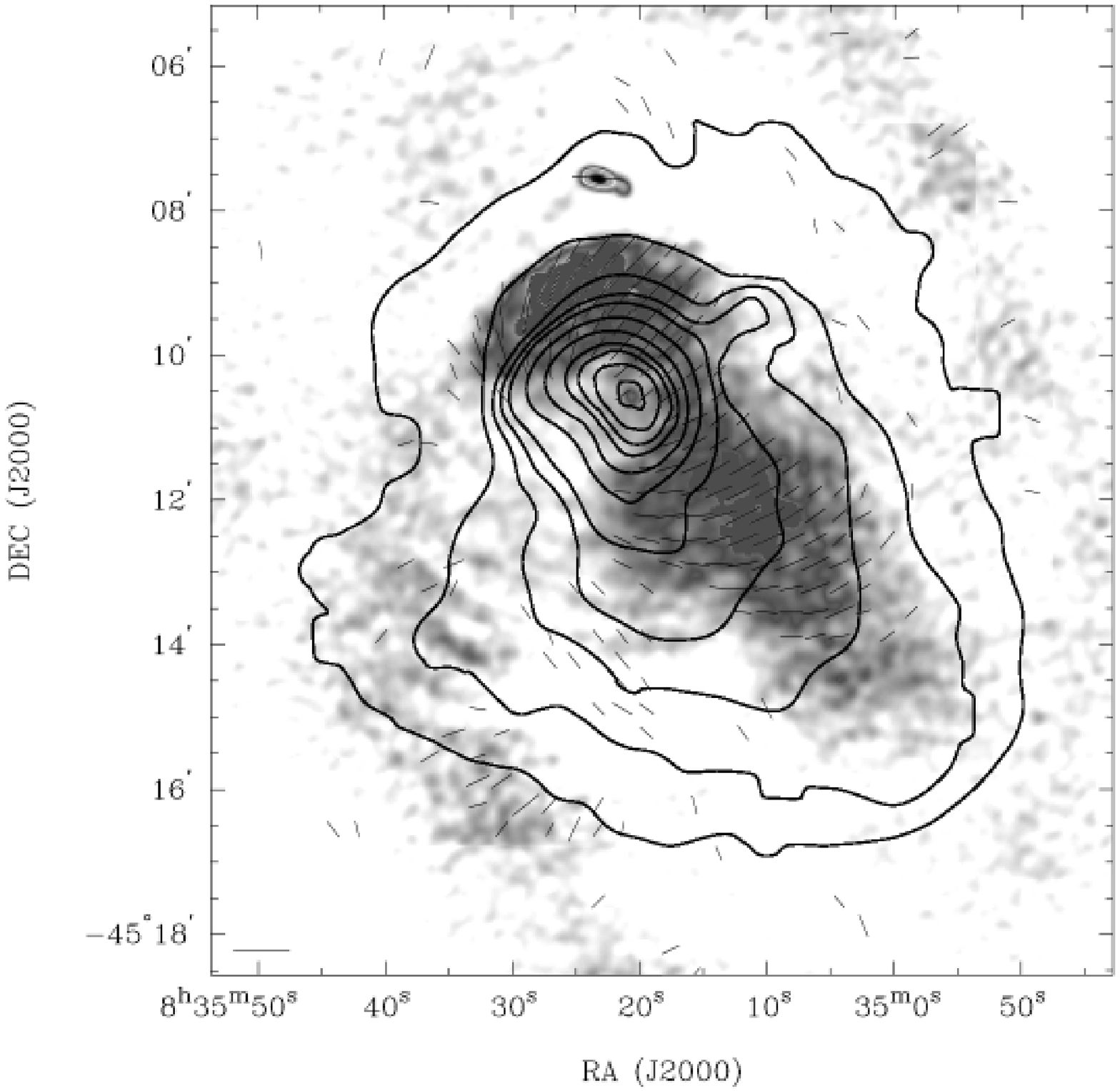,height=6.0cm} \hspace{0.1cm}
\psfig{figure=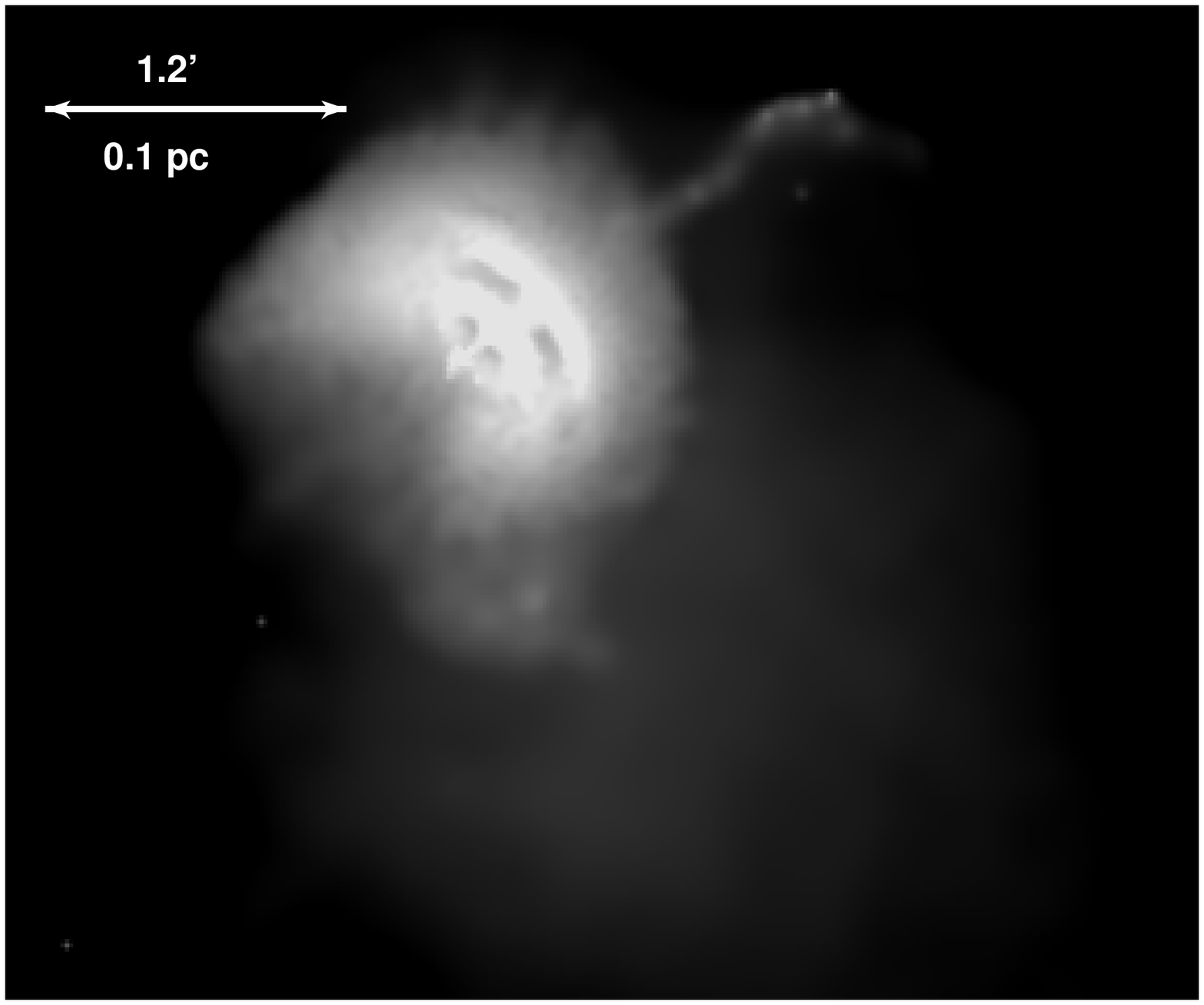,height=6.0cm}
 }
\vspace{-0.8cm}
\end{center}\caption{ {\bf Left:} 8.5 GHz image (from Dodson et al. 2003b)
with the X-ray contours overlayed. {\bf Right:} Smoothed version
of the summed image of the Vela PWN showing the structure of the
inner PWN. }
 \label{fig2} \end{figure}

To conclude, the multiwavelength data continue to provide clues to
understanding the complicated PWN topology and elucidate the
dynamics of the relativistic pulsar winds and their interaction
with the SNR medium.

\acknowledgments
Support for this work was provided by NASA
through grant NAG5-10865 and {\sl Chandra} Award GO2-3091X.

\end{document}